\documentclass[aps, showpacs, reprint, twocolumn, superscriptaddress]{revtex4-1}
\usepackage{bm}
\usepackage{graphicx}
\usepackage{units}
\usepackage{amssymb,amsmath,amsthm,color}
\usepackage{microtype}

\renewcommand{\vec}[1]{\bm{#1}}
\newcommand{\uvec}[1]{\bm{\hat{#1}}}
\newcommand{\avr}[1]{\left\langle#1\right\rangle}

\DeclareMathOperator{\erfc}{erfc}

\begin{document}
\title{Effective dynamics of microorganisms that interact with their own trail}

\author{W. Till Kranz}
\affiliation{Rudolf Peierls Centre for Theoretical Physics, University of Oxford, Oxford OX1 3NP, United Kingdom}

\author{Anatolij Gelimson}
\affiliation{Rudolf Peierls Centre for Theoretical Physics, University of Oxford, Oxford OX1 3NP, United Kingdom}

\author{Kun Zhao}
\affiliation{Key Laboratory of Systems Bioengineering, Ministry of Education, School of Chemical Engineering and Technology, Tianjin University, Tianjin, 300072, People's Republic of China}
\affiliation{Bioengineering Department, Chemistry \& Biochemistry Department, California Nano Systems Institute, UCLA, 90095-1600, Los Angeles, CA, USA}

\author{Gerard C. L. Wong}
\affiliation{Bioengineering Department, Chemistry \& Biochemistry Department, California Nano Systems Institute, UCLA, 90095-1600, Los Angeles, CA, USA}

\author{Ramin Golestanian}
\email{ramin.golestanian@physics.ox.ac.uk}
\affiliation{Rudolf Peierls Centre for Theoretical Physics, University of Oxford, Oxford OX1 3NP, United Kingdom}

\date{\today}

\begin{abstract}
  Like ants, some microorganisms are known to leave trails on surfaces
  to communicate. We explore how trail-mediated self-interaction could
  affect the behavior of individual microorganisms when diffusive
  spreading of the trail is negligible on the timescale of the
  microorganism using a simple phenomenological model for an actively
  moving particle and a finite-width trail. The effective dynamics of
  each microorganism takes on the form of a stochastic integral
  equation with the trail interaction appearing in the form of
  short-term memory. For moderate coupling strength below an emergent
  critical value, the dynamics exhibits effective diffusion in both
  orientation and position after a phase of superdiffusive
  reorientation. We report experimental verification of a seemingly
  counterintuitive perpendicular alignment mechanism that emerges
  from the model.
\end{abstract}

\pacs{87.18.Gh, 87.17.Jj, 87.10.Ca}

\maketitle

For many animals and microorganisms it is advantageous to know where
their companions or they themselves have been
\cite{couzin09,passino+seeley08,reid+latty12,benjacob+cohen00,chowdhury+nishinari04,jackson+martin06,bonner+savage47,kaiser+crosby83,wu+jiang07}. To
this end, many creatures leave trails of some characteristic
substance. A well studied example is the pheromone trails of ants
\cite{jackson+martin06,sumpter+beekman03}, which allows them to
collect food efficiently. Single cell organisms are also known to
leave trails \cite{burchard82,zhao+tseng13}. It is believed that the
trails help them form aggregates in sparse populations
\cite{bonner+savage47,kaiser07,zhao+tseng13}, whereas in denser
populations, colonies could also result from the combined effect of
surface-bound motility and excluded volume interactions
\cite{peruani+starruss12,soto+golestanian14}. For bacteria, these
trails are often subsumed as exopolysaccharides (EPS)
\cite{sutherland01} but may also contain proteins
\cite{dworkin+kaiser93}. To be evolutionarily favorable, the
(energetic) costs incurred by trail formation should balance the
advantages gained through this form of communication.

Chemotaxis is commonly mediated by rapidly diffusing signalling
molecules \cite{bonner+savage47} but, more generally, cell-cell
signalling can also be mediated by trails of macromolecules that
diffuse much more slowly than the microorganism or form stable gels
\cite{sutherland01,christensen+characklis90}. Chemically-mediated
interactions between bacteria or eukaryotic cells
\cite{brenner+levitov98,benjacob+cohen00,tsori+degennes04,taktikos+zaburdaev11,taktikos+zaburdaev12,levine+rappel13,gelimson+golestanian15}
as well as artificial active colloids
\cite{golestanian12,cohen+golestanian14,saha+golestanian14} lead to a
variety of collective phenomena including collapse, pattern formation,
alignment and oscillations. Auto-chemotactic effects have
  been studied in the context of swimming bacteria
\cite{grima05,sengupta+vanteeffelen09} and {\em Dictyostelium} cells
\cite{westendorf+negrete13}. While much is known about the chemotactic
machinery in bacteria \cite{sourjik+berg04,tu+grinstein05} and
eukaryotic cells \cite{levine+rappel13}, relatively little is known
about trail-mediated interactions.

Whereas ants have antennae that are spatially well separated from
their pheromone glands \cite{hangartner67}, such a clear separation is
difficult for single-celled organisms
\cite{dworkin+kaiser93,thar+kuehl03,herzmark+campbell07}. In addition
to sensing the trails left by other individuals, microorganisms are
also immediately affected by their own trails. This suggests that
trail-mediated self-interaction could play a significant role in the
behavior of microorganisms. For example, by providing a mechanism to
tune the effective translational and orientational diffusivities, or
by creating distinct modes of motility, and consequently, the search
strategy.

\begin{figure}[h]
  \centering
  \includegraphics{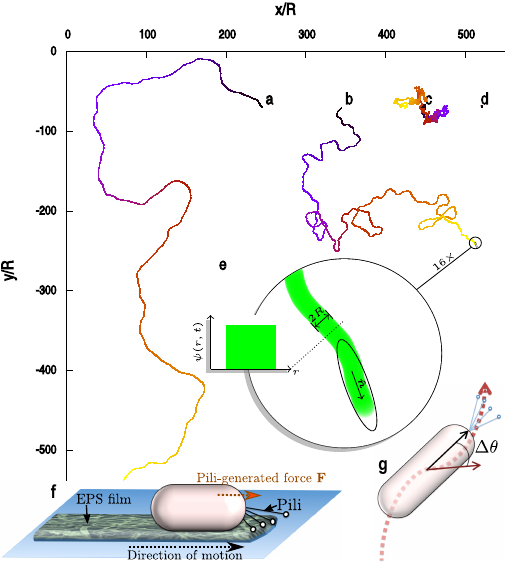}
  \caption{(color online). Sample trajectories generated by the
    effective dynamics, Eqs.~(\ref{eq:1},\ref{eq:3},\ref{eq:4}), over
    a period of time $10^3\tau$ (color coded) for: no interaction with
    the trail, $\Omega\tau\equiv0$ (a), weak interaction,
    $\Omega\tau=1.2$ (b), strong interaction, $\Omega\tau = 1.85$ (c),
    close to the localization transition, and above it,
    $\Omega\tau=2.15$ (d). The rotational diffusivity is set to
    $D_r^0\tau = 10^{-2}$, and $2\tau$ is the trail crossing
    time. Panel (e) shows a magnification of the end of trail b with
    the trail field, $\psi(\vec r, t)$, of width $2R$ in green and the
    current orientation of the microorganism (ellipse), $\uvec n$. (f)
    Schematic depiction of a microscopic model system such as \textit
    {P. aeruginosa} that uses pili for motility and sensing.
    (g) A schematic for the definition of $\Delta \theta$, which is
    the angle between the current body orientation and the bacterial
    trajectory (dotted line).}
  \label{fig:intro}
\end{figure}

\begin{figure*}
  \centering
  \includegraphics{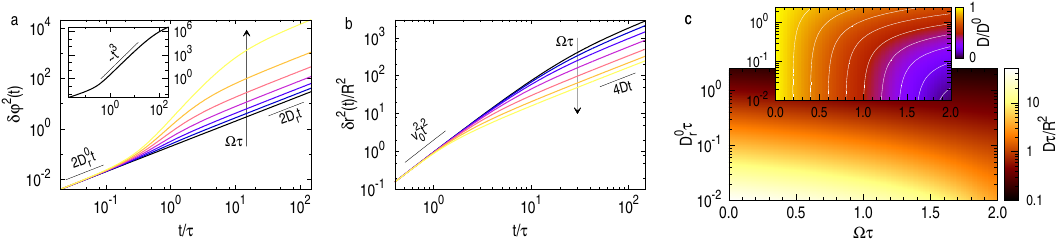}
  \caption{(color online). Angular \textsc{MSD} $\delta\varphi^2(t)$
    (a), and translational \textsc{MSD} $\delta r^2(t)$ normalized by
    the trail width, $R$, (b) as a function of time $t$ for several
    values of the effective turning rate
    $\Omega\tau = 0, 0.4, 0.7, 1.0, 1.3, 1.6, 1.9$. The
    microscopic diffusivity is set to $D_r^0\tau=0.1$. The inset of (a) shows $`\delta\varphi^2(t)$ for $\Omega\tau=1.99$ demonstrating the intermediate superballistic regime. (c)
    Color coded translational diffusivity $D$ normalized by the trail width $R$
    and the trail crossing time $\tau$ as a function of the control
    parameters. Note the logarithmic axes. \emph{Inset:} The
    translational diffusivity for the same range of parameters
    normalized by the trail free ($\Omega\tau\equiv0$) value $D^0$ on
    a linear scale. The white contour lines are $0.1$ apart.
    \label{fig:2}
  }
\end{figure*}

In this Letter, we discuss a simple but generic model of a
microorganism experiencing trail-mediated interactions. Focusing on a
persistent EPS trail with vanishing diffusivity but taking its
finite width explicitly into account, we focus on the immediate
self-interaction that previously had to be excluded \textit{a priori}
by an \textit{ad hoc} refractory period
\cite{tenhagen+vanteffelen11,taktikos+zaburdaev12}. While previous
work has mostly considered a concentration dependent speed
\cite{tsori+degennes04,grima05,sengupta+vanteeffelen09}, a coupling to
the orientation arises naturally \cite{taktikos+zaburdaev11,saha+golestanian14,Note2}. We
find that the self-trail interaction modifies the translational and
orientational motion of the microorganisms and renormalize the
corresponding diffusion coefficients, at the longest time scale (see
Fig.~\ref{fig:intro}).

\paragraph*{Microscopic Model.---}
\label{sec:microscopic-model}

We consider a single particle of width $2R$ whose state at
time $t$ is defined by its position $\vec r(t)$ and orientation $\uvec
n(t) = (\cos\varphi, \sin\varphi)$. We model the dynamics by
prescribing a fixed characteristic speed, $v_0$, for the particle,
namely
\begin{subequations}
  \begin{equation}
    \label{eq:1}
    \partial_t\vec r(t) = v_0\uvec n(t).
  \end{equation}
  The motion will typically be generated via the cooperation of a
  number of molecular motors, whether it is realized by the retraction
  of pili \cite{skerker+berg01,holz+opitz10}, the extrusion of slime
  \cite{wolgemuth+hoiczyk02}, or any other mechanism. This implies
  significant noise in the propulsion force, and consequentially, a
  finite directional persistence. For the simplest, trail-free case,
  we model the orientational dynamics as a purely diffusive process,
  $\partial_t\varphi(t) = \xi(t)$, where $\xi(t)$ is a Gaussian random
  variable obeying $\avr{\xi(t)\xi(t')} = 2D_r^0 \delta(t - t')$ and
  $D_r^0$ is the microscopic rotational diffusion coefficient
  controlling the persistence time $1/D_r^0$. This trail-free model
  displays a translational mean-square displacement (\textsc{MSD})
  $\delta r^2(t) = \avr{[\vec r(t)-\vec r(0)]^2}$ that crosses over
  from ballistic, $\delta r^2(t) = v_0^2t^2$, for, $D_r^0t\ll 1$, to
  diffusive behavior, $\delta r^2(t) = 4 D^0 t$, for $D_r^0 t \gg 1$
  where $D^0 = v_0^2/(2 D_r^0)$. Fluctuations in $v_0$ could also be
  taken into account in a straightforward generalization
  \cite{peruani+morelli2007}.

  The trail excreted from the microorganism can be characterized by
  the density profile $\psi(\vec r, t)$ that satisfies the diffusion
  equation $\partial_t \psi(\vec r, t)-{\cal D}_{\rm p} \nabla^2
  \psi(\vec r, t)=k \delta_R^2\left(\vec r - \vec r(t)\right)$, where
  $k$ is the deposition rate and $\delta_R^2\left(\vec r - \vec
    r(t)\right)$ is a ``regularized delta function'' that accounts for
  the finite size $R$, and traces
  its position [normalized as $\int d^2r\,\delta_R^2(\vec r) =
  1$]. Setting ${\cal D}_{\rm p}=0$, we find for the trail profile at
  time $t$ and position $\vec x$ as
\begin{equation}
  \label{eq:2}
  \psi(\vec x, t) = k\int_0^t d t'\,\delta_R^2\left(\vec x - \vec r(t')\right).
\end{equation}
We choose a rectangular source, $\delta_R^2(\vec r) = \Theta(R^2 -
r^2)/\pi R^2$, where $\Theta(x)$ denotes the Heaviside step function
and $r\equiv|\vec r|$. The trail width $2R$ defines a
microscopic time scale $\tau = R/v_0$, which gives the trail crossing
time (see Fig.~\ref{fig:intro}e). This specific regularization scheme
is a good representation of the regime in which the characteristic
diffusion length of the polymeric trail is much smaller than the width
of the trail, $\sqrt{{\cal D}_{\rm p} \tau} \ll R$ \cite{Note1}.

A generic interaction with the trail couples to gradients of the
trail field perpendicular to the current orientation
\cite{taktikos+zaburdaev12,saha+golestanian14}, effectively steering
the microorganism toward trails by favouring a orientation $\uvec n$
perpendicular to the trail, i.e.,
\begin{equation}
  \label{eq:3}
  \partial_t\varphi(t) = \chi\partial_{\perp} \psi(\vec r(t), t) + \xi(t),
\end{equation}
where $\partial_{\perp} \psi = \uvec n_{\perp}(t)\cdot\nabla\psi(\vec r(t), t)$, with $\uvec n_{\perp}
= (-\sin\varphi,\cos\varphi)$ being the angular unit vector in polar
coordinates. The sensitivity to the trail is controlled by a parameter
$\chi$. We have provided a microscopic derivation of this coupling
\cite{Note2} for a model system of a pili-driven bacterium on a
substrate (see Fig.~\ref{fig:intro}f) by assuming a generic dependence
of the pili surface attachment force on the EPS concentration.
However, Eq. (\ref{eq:3}) will be expected in the continuum limit for
any microscopic model based on symmetry considerations
\cite{saha+golestanian14}.

\paragraph*{Effective Dynamics.---}
\label{sec:effective-dynamics}

We assume that the particle trajectory does not bend back on itself
immediately (no-small-loops assumption), and that self-intersections
on longer times are rare enough to be negligible.

By making a short time expansion of Eqs.~(\ref{eq:1}) and (\ref{eq:3})
to be inserted into Eq.~(\ref{eq:2}), one finds a closed equation for
the head of the trail field $\partial_{\perp} \psi(t)
\equiv \partial_{\perp} \psi(\vec r(t), t)$ \cite{Note2}. The result
is a stochastic integral equation
  \begin{equation}
    \label{eq:4}
    \partial_{\perp} \psi(t) = \frac{\Omega}{\tau}\int_0^{\tau} d u \, (\tau-u) \,
    [\partial_{\perp} \psi(t-u) + \xi(t- u)/\chi],
  \end{equation}
\end{subequations}
The effective turning rate $\Omega = k\chi\tau/\pi R^3$ increases for
more intense trails (larger $k$) and for more sensitive organisms
(larger $\chi$). The delay $\tau$ reflects the memory imparted by the
trail. The closed set of equations (\ref{eq:1},\ref{eq:3},\ref{eq:4})
constitute our effective dynamical description of the system.

\begin{figure}[tb]
  \centering
  \includegraphics{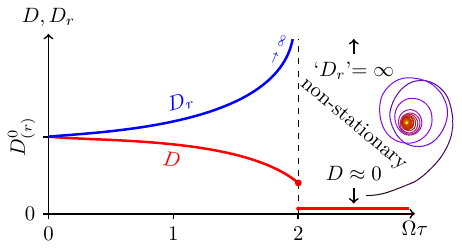}
  \caption{(color online). Phase diagram of the dynamics of the
    microorganism with trail-mediated self-interaction, as a function
    of the dimensionless turning frequency $\Omega
    \tau$. \textit{Inset:} Zoomed view of Fig.~1d.}
  \label{fig:phase}
\end{figure}

For the average gradient one finds
$\avr{\partial_{\perp}\psi}\sim\exp(\alpha t)$ where the rate $\alpha$
is given implicitly as the solution of $\lambda(\alpha) = 0$ where
$\lambda(\alpha) = 1 - \frac{\Omega\tau}{\alpha\tau}
\left[1 + \frac{1}{\alpha\tau}\left(e^{-\alpha\tau} - 1\right)\right]$.
For $\Omega\tau < 2$, $\alpha < 0$ and Eq.~(\ref{eq:4}) defines a
random process with zero mean that leads to a stationary dynamics
which is time-translation invariant.  For $\Omega\tau > 2$ one finds
$\alpha>0$, i.e., the gradient (angular velocity) diverges
exponentially in time. Implying that the trajectory converges in a
logarithmic spiral to a localized point. This means there is a maximum
value of the product of trail deposition rate and sensitivity,
$k\chi$, that allows steady-state motion. For sample trajectories, see
Figs.~\ref{fig:intro} and \ref{fig:phase}.

In the stationary regime, Eq.~(\ref{eq:4}) can be solved in the
frequency domain, $\chi\widetilde{\partial_{\perp}\psi}(\omega) =
[\lambda^{-1}(i\omega) - 1]\tilde\xi(\omega)$. The trail-mediated
self-interaction thus linearly transforms the intrinsic white noise
$\xi$ to an effective colored random angular velocity,
$\tilde\Xi(\omega) = \tilde\xi(\omega)/\lambda(i\omega)$ such that
$\partial_t\varphi(t) = \Xi(t)$.

\paragraph*{Angular \& Translational MSD.---}
\label{sec:msd}

The most easily accessible quantity in experiments is the
translational \textsc{MSD} $\delta r^2(t)$, which is related to the
angular \textsc{MSD}, $\delta\varphi^2(t)= \avr{[\varphi(t) - \varphi(0)]^2}$, via \cite{doi+edwards88}
\begin{equation}
  \label{eq:6}
  \delta r^2(t) = 2v_0^2\int_0^tdt' \, (t-t') \, e^{-\delta\varphi^2(t')/2}.
\end{equation}
The angular \textsc{MSD} is a sum of three terms, given in the Laplace
domain as
\begin{math}
  s^2\widehat{\delta\varphi^2}(s)/(2D_r^0) = 1 + \widehat{\Delta}(s)
  + \widehat\Lambda(s).
\end{math}
The corrections to simple diffusion are given by the two correlation
functions $\Lambda(t) := \chi\avr{\partial_{\perp} \psi(t)
  \xi(0)}/(2D_r^0)$ (such that $\hat\Lambda(s) = \lambda^{-1}(s) - 1$)
and $\Delta(t)= \chi^2\avr{\partial_{\perp} \psi(t)\partial_{\perp} \psi(0)}/D_r^0 =
2\int_0^tdt'\Lambda(t')\Lambda(t' + t)$.
The definition of $\hat\Lambda(s)$ shows that $\Omega\tau$ is the only
relevant control parameter for the orientational \textsc{MSD}
$\delta\varphi^2(t)$ and the behavior of $\delta\varphi^2(t)$ is
fully determined by the analytic structure of $\lambda^{-1}(s)$.

In the stationary regime, $\delta\varphi^2(t)$ (cf.~Fig.~\ref{fig:2}a
\cite{Note3}) starts off
diffusively, $\delta\varphi^2(t) = 2D_r^0t$ for $t \ll \tau$ and
becomes asymptotically diffusive again, $\delta\varphi^2(t) = 2D_rt$
for $t\gg\tau/(1 - \Omega\tau/2)$ determined by the smallest
  pole of $\lambda^{-1}(i\omega)$.  The effective orientational
diffusivity
\begin{equation}
  \label{eq:8}
  D_r/D_r^0 = 1 + \frac{\Omega\tau}{2}
  \times\frac{1 + \Omega\tau/2}{(1 - \Omega\tau/2)^2},
\end{equation}
diverges for $\Omega\tau\to2$ confirming our expectation that the
trail-mediated self-interaction reduces orientational persistence. The
two diffusive regimes are joined by an intermediate, superdiffusive
regime. Note that the crossover time to the asymptotic diffusive
regime diverges as $\Omega\tau\to2$. Close to the limiting value
$\Omega\tau = 2$, the crossover is given by the super-ballistic law
$\delta\varphi^2(t) = 6D_r^0\tau(t/\tau)^3$ for $\tau \ll t \ll
\tau/(1 - \Omega\tau/2)$. For a fast effective turning rate $\Omega >
D_r^0$, the intrinsic noise combines a diffusive ($\propto
t^{1/2}$) excursion with the ballistic ($\propto t$) reorientation
due to the self-interaction, leading to $\delta\varphi(t) \propto
t^{3/2}$ until later times where the stochastic character of the
self-interaction becomes important and turns the behavior back to diffusion.

The translational \textsc{MSD} $\delta r^2(t)$ always starts ballistically,
$\delta r^2(t) = v_0^2t^2$ for $t \ll \tau$ and crosses over to diffusive
behavior $\delta r^2(t) = 4Dt$ for long times $t\to\infty$
(cf.~Fig.~\ref{fig:2}b). The crossover time, $t^*$ will be determined
implicitly by $\delta\varphi^2(t^*) \sim 1$. For the location of the crossover
and the dependence of the translational diffusivity on the control parameters,
we have to consider a number of different regimes.

\paragraph*{Short Persistence Regime.---}
\label{sec:short-persistence}

When $(D_r^0\tau)^{-1} < 1 + \Omega\tau/2$, the crossover happens around
$\Omega t \sim \sqrt{1 + 2\Omega/D_r^0} - 1$ and the asymptotic diffusivity
\begin{equation}
  \label{eq:12}
  D/D^0 = \sqrt{\pi
    D_r^0/2\Omega}\,e^{D_r^0/2\Omega}\erfc(\sqrt{D_r^0/2\Omega}),
\end{equation}
is a function of the ratio $D_r^0/\Omega$ alone.

\paragraph*{Long Persistence Regime.---}
\label{sec:long-pers-regime}

For sufficiently straight trails such that by the time $\delta\varphi^2(t)
\sim 1$ it is already deep in the long time diffusive regime, i.e.,
$D_r^0\tau \ll 2(1 - \Omega\tau/2)^3/[2 - \Omega\tau +
(\Omega\tau)^2/2]$, the crossover happens around $t \sim 1/D_r$ and
the asymptotic diffusivity is given as
\begin{equation}
  \label{eq:13}
  D/D^0 = \frac{D_r^0}{D_r}\left[
    1 - \frac{(D_r^0\tau)^2}{6}
    \frac{\Omega\tau(1 + \Omega\tau/2)^3}{(1 - \Omega\tau/2)^6}
  \right].
\end{equation}

\paragraph*{Critical Regime.---}
\label{sec:critical-regime}

Close to the upper limit $\Omega\tau\to2$ and for $D_r^0\tau < 1$, the
crossover happens around $t/\tau \sim 1/\sqrt[3]{D_r^0\tau}$ and the
asymptotic diffusivity
\begin{equation}
  \label{eq:14}
  D/D^0 = \Gamma(3/4)(D_r^0\tau)^{2/3} + D_r^0\tau/3,
\end{equation}
is a function of $D_r^0\tau$ alone. Note that the dependence on the
intrinsic noise, $D \propto 1/\sqrt[3]{D_r^0}$, is significantly
weakened compared to the trail free case, $D^0\propto
1/D_r^0$, and that $D$ does not vanish as $\Omega\tau\to2$.

\paragraph*{Intermediate Regime.---}
\label{sec:intermediate-regime}

In the rest of the parameter space of $D_r^0\tau$ and $\Omega\tau$, no
explicit expressions can be given and the asymptotic diffusivity $D$
will be a function of both control parameters. Numerical result for
the effective translational diffusivity is presented in
Fig.~\ref{fig:2}c.

\paragraph*{Discussion.---}
\label{sec:discussion}

The interaction of a microorganism with its own trail effectively
introduces a new timescale $1/\Omega$. The trail-mediated
self-interaction modulates the intrinsic noise in a linear but
nontrivial way. While the asymptotic dynamics remains diffusive below
the critical value $\Omega\tau = 2$, both for the translational and
the orientational degrees of freedom, the (orientational)
diffusive regime may only be reached on timescales that may be much
longer than $\tau$. This holds in the limit of strong interactions
where the crossover timescale $\tau/(1 - \Omega\tau/2)$ becomes
increasingly large but also for large effective persistence times
$1/D_r$. Moreover, the crossover times are distinct from the
microscopic persistence time $1/D_r^0$.

The asymptotic angular diffusivity, $D_r$, is a function of
$\Omega\tau$ that diverges as $\Omega\tau\to2$. It is always larger
than its microscopic value $D_r^0$, i.e., orientational persistence is
reduced by the trail. The translational diffusivity, $D$, on the other
hand, is always reduced and, in general, depends on the parameters
$\Omega\tau$ and $D_r^0\tau$.

For very strong trail-mediated self-interactions, $\Omega\tau > 2$,
the initial dynamics quickly confines the particle in a region of size
$\lesssim 2R$. Here, our assumptions break down and the ensuing
dynamics will depend on more microscopic details not resolved in the
present model. Nevertheless, we can formally identify the behavior of
the microorganism by a diverging rotational diffusivity. The
translational diffusivity incurs a finite jump at the transition point
as it is determined by the regular short-time part of
$\delta\varphi^2(t)$ below the transition and by unresolved
microscopic details above it. We note that the no-small-loops
condition, which can be written as $\delta\varphi^2(\tau) \sim (\Omega
\tau)^2 (D_r^0\tau)^2 < \pi^2$, does not obscure this phase
transition, so long as $D_r^0 \tau < 1$. The analysis of a more
microscopic model is beyond the scope of this contribution. The
behavior of the system can be summarized in a phase diagram as shown
in Fig.~\ref{fig:phase}. Our findings are subtly different from the
sub-diffusion observed in Ref.~\cite{chepizhko+peruani13} as a result
of temporary trapping of bacteria in loops caused by quenched
disorder, which is a stationary regime.

\begin{figure}[tb]
  \centering
  \includegraphics[width=\columnwidth]{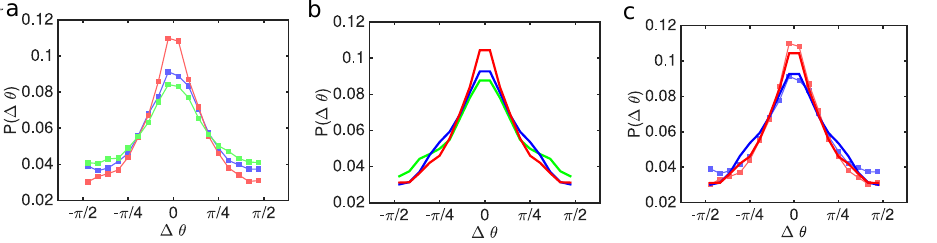}
  \caption{(color online). (a) The experimentally observed narrowing
  of the $\Delta \theta$ distribution with increasing trail deposition
  (see Fig. \ref{fig:intro}g for the definition of $\Delta \theta$).
  The experiments were carried out with the \textit{P. aeruginosa}
  mutants $\Delta psl$, which does not secrete Psl (light green) and
  $\Delta$P$_{psl}$/P$_{BAD}$-{\it psl}, which secretes Psl in response to the arabinose
  in the environment. For $\Delta$P$_{psl}$/P$_{BAD}$-{\it psl}, the arabinose
  concentration was varied between $0 \%$ (light blue, low Psl deposition)
  and $1 \%$ (light red, high Psl deposition). (b) The corresponding
  theoretical $\Delta \theta$ distributions resulting from the influence
  of the alignment term. The trail coupling strength is varied between
  $\Omega \tau = 0.05$ (green) $\Omega \tau = 0.12$ (blue) and
  $\Omega \tau = 0.35$ (red). The distribution has been sampled
  using the time interval of $\Delta t=0.2 \, \tau$, and the rotational
  diffusion has been set to $D_r^0 \tau=0.015$. These values roughly
  correspond to the experimental parameters at which the distribution
  has been measured. (c) A comparison between experimental (dots)
  and theoretical  (lines) distributions of $\Delta \theta$ for
  the $\Delta$P$_{psl}$/P$_{BAD}$-{\it psl} mutant under $0 \%$ (blue) and $1 \%$ arabinose (red).
  }
  \label{dtheta-distribution}
\end{figure}

The perpendicular alignment strategy might appear to be counter-intuitive,
but it is supported by recent experimental evidence. We have investigated
this question by probing the angle distributions of single
bacteria in experimentally recorded motion of \textit{P. aeruginosa}
with different rates of Psl exopolysaccharide deposition (in a similar
experiment to Ref. \cite{zhao+tseng13}); for methods, see the Supplemental Material \cite{Note2}.
For the angle $\Delta\theta$ between the trajectory and the body orientation
(defined in Fig. \ref{fig:intro}g), the experiments show that the distribution
narrows down with increasing increasing the secretion of Psl trails in via a
mutant with an arabinose inducible promoter, which increases the effective
trail deposition strength; see Figs. \ref{dtheta-distribution}b and c.
Our numerical simulations of the model predict a similar trend, as shown in
Fig. \ref{dtheta-distribution}b and c, highlighting the interplay between the noise
and the tendency of perpendicular alignment. The trail can have a
stabilizing effect by reorienting the microorganism towards the inner regions
in case it reaches the trail boundaries. The result is a significantly narrower
$\Delta \theta$ distribution with increasing $\Omega$, indicating a higher
correlation between trajectory and microorganism orientation. Further evidence
in support of the perpendicular alignment scenario can be extracted from
the collective behavior of the bacteria \cite{New-Gelimson-etal}.

Our results could have significant biological (as well as
biophysical \cite{Note2}) implications. Regulating the strength of
the trail-mediated self-interaction may allow microorganisms
to decide whether to confine themselves to smaller areas and
search them more thoroughly or explore larger areas.
Interestingly, the effective translational diffusion
coefficient scales as $1/\sqrt[3]{D_r^0}$ in the presence of strong
trail-mediated self-interaction, as compared to $1/D_r^0$ in the
trail-free case. This suggests that trails make the microorganism
less sensitive to intrinsic variations in the orientational noise.

To conclude, we have shown that a very simple model of particle-trail
interaction leads to a wealth of nontrivial phenomena, including an
transition from stationary to non-stationary behavior with a
diverging orientational diffusivity. Our results could shed light on
the behavior of trail-forming microorganisms, and in particular how
they can use this ``expensive'' output to regulate their own activity
while, simultaneously, providing a communication channel with other
individuals. Moreover, they could also find use in the field of
robotics by providing a blue-print for designing micro-robots
that can tune their search strategy via local interactions
with their own trails.

\begin{acknowledgments}
  We thank Ben Hambly for insightful discussions and Tyler Shendruk
  for carefully reading the manuscript. This work was supported
  by the Human Frontier Science Program RGP0061/2013.
\end{acknowledgments}

%

\appendix

\section{Microscopic derivation of the equations of motion}

\begin{figure}
  \centering
  \includegraphics[width=0.48\textwidth]{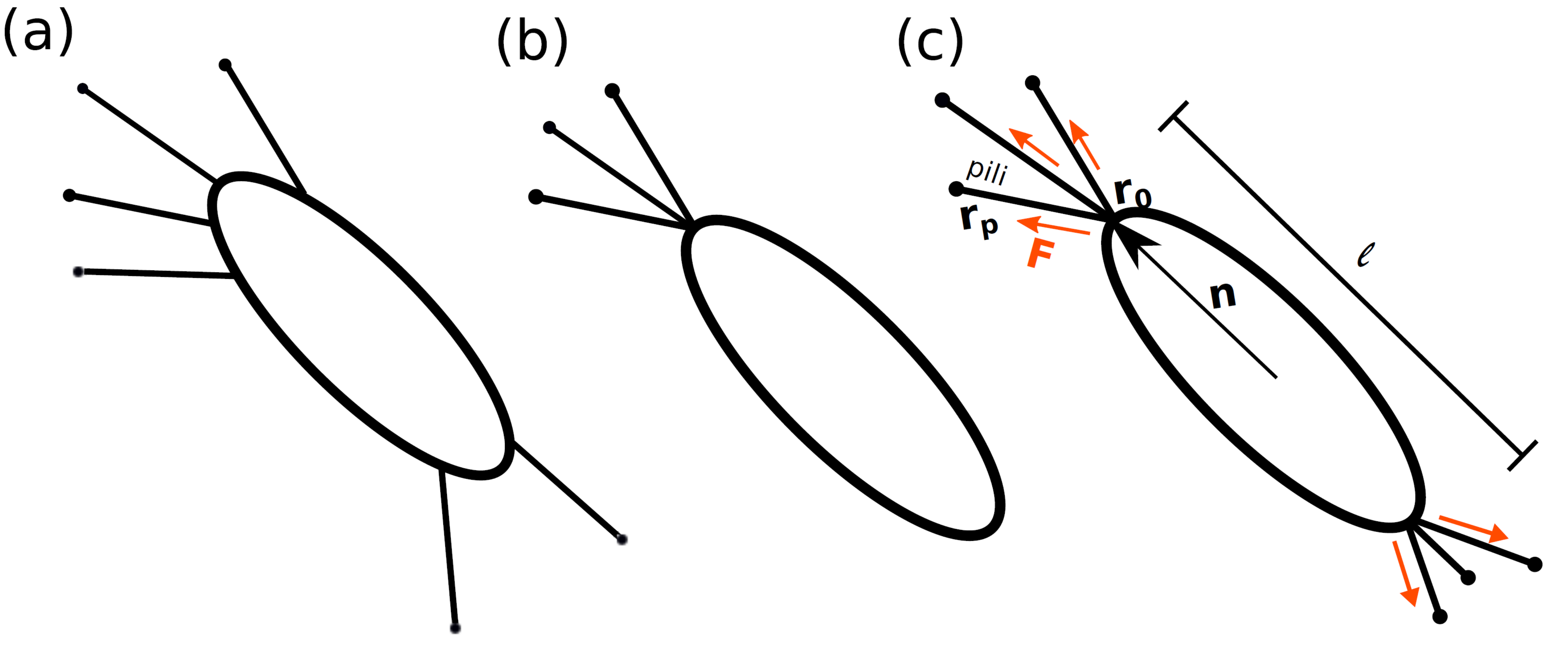}
  \caption{Schematics of pili distribution around a bacterium. (a)
    Asymmetric distribution around the body (\textit{N. gonorrhoae}).
    (b) Pili concentrated at one pole (\textit{P. aeruginosa}). (c)
    Idealized model with a symmetric distribution of pili at the
    poles.}
  \label{fig:model_simple}
\end{figure}

We regard a bacterium of length $\ell$ which secretes an EPS film and pulls
itself forward using typically $N$ pili, Fig.~\ref{fig:model_simple}. We will
assume that the pili attach to the surface in an EPS-dependent way. Better
attachment results in more effective pulling and therefore we can assume that
the pulling force of an individual pilus attached at site $\vec r_i$ is
EPS-dependent, $f = f(\psi(\vec r_i))$. The distribution of pili on the
surface of bacteria varies among species. They may be scattered all over the
surface (Fig.~1a), or concentrated in one (Fig.~1b), or both poles (Fig.~1c)
\cite{maier13}. The general argument given below holds for all these pilus
configurations but to be concrete, we will assume a symmetric distribution of
an equal number of pili, concentrated at the two poles. The attachment sites
are assumed to be distributed randomly according to some distribution $P(\vec
r_i - \vec r_0; \uvec n) = P(\vec r_p; \uvec n) = P(r_p, \vartheta)$, where
$\vec r_0$ is the head (or tail) of a bacterium, $\vec r_p$ is the relative
coordinate of the pilus tip, and $\vartheta$ is the angle between $\uvec n$
and $\vec r_p$. We assume that the force $\vec f_i = f(\psi(\vec r_i))\uvec
e_p$ will pull along the direction of $\vec r_p$. It is plausible to assume
that the probability for directions in which pili can face is symmetrically
distributed around the body orientation $\uvec n$. This implies that $\int
\uvec e_p P(\vec r_p; \uvec n) d^2r_p \parallel \uvec n$. If this was not the
case, the pili would preferentially explore the space right/left of the body
and there would be, to a first approximation, a constant torque turning the
microorganism in one direction.

We assume that $\psi(\vec r)$ is a smooth function on the scale of a
typical pilus length $\langle r_p\rangle$ and can be expanded as
\begin{equation}
  f(\psi(\vec r_i)) = f(\psi(\vec r_0))
  + f'(\psi(\vec r_0))[\nabla\psi(\vec r_0)\cdot\vec r_p]
  + \mathcal O(\vec r_p^2)\notag
\end{equation}
where $f' \equiv \partial_{\psi}f(\psi)$.

We can then use this to approximate the average total force
$\langle\vec F\rangle = \langle\sum_i\vec f_i\rangle$
\begin{equation}
  \begin{aligned}
    \langle\vec F\rangle \approx &N\int d^2r_pP(\vec r_p; \uvec n)f(\psi(\vec r_0 + \vec r_p))\uvec e_p\\
    \approx &F(\psi(\vec r_0))\uvec n
    \int_{-\pi}^{\pi}\!\!\cos\vartheta P(\vartheta)d\vartheta\\
    &+ F'(\psi(\vec r_0))\nabla\psi(\vec r_0)\cdot\int d^2r_pP(\vec r_p; \uvec n)
    r_p\uvec e_p\uvec e_p
    \label{eq:force-expansion}
  \end{aligned}\notag
\end{equation}

The pili force component parallel to the bacterial body will leave
the orientation of the bacterium unchanged and propel the centre of
mass of the bacterium. The average propelling force will be
\begin{equation}
\langle \vec F_{\parallel} \rangle = \uvec n(\uvec n\cdot\langle\vec F\rangle)
\approx \uvec n F(\psi(\vec r_0))\langle\cos\vartheta\rangle
\label{eq:pullingforce-average}
\end{equation}
In an overdamped system, the velocity of the microorganism will be
proportional to the pulling force $\partial_t\vec r =
\mu_{\parallel}\langle\vec F_{\parallel}\rangle = v_0\uvec n$ where
$\mu_{\parallel}$ is the translational mobility. This gives equation
(1a) with
\begin{equation}
  v_0(\psi) = \mu_{\parallel}F(\psi(\vec r))\langle\cos\vartheta\rangle
\end{equation}
The velocity of the microorganism is in general $\psi$-dependent but
in case of a large constant contribution of $\vec F$, the
$\psi$-dependent contribution to the velocity will be subdominant.

The perpendicular component of the pulling force $\vec F_{\perp} =
\vec F - \uvec n(\uvec n\cdot\vec F)$, on the other hand, will
generate a torque on the body of the microorganism and here, the EPS
dependence of the force needs to be taken into account even at the
lowest order. The average torque is given by $\langle\vec\tau\rangle =
(\ell/2)\uvec n\times\langle\vec F\rangle$. Using
Eq.~(\ref{eq:force-expansion}) we get
\begin{equation}
  \langle\vec\tau\rangle \approx \frac\ell2F'(\psi(\vec r_0))
  [\uvec n\times\nabla\psi(\vec r_0)]\langle r_p\sin^2\vartheta\rangle
\end{equation}
and a completely equivalent equation for the torque coming from pili
pulling at the other tip of the microorganism.

On a surface the motion of the microorganism will be overdamped and
therefore the angular velocity $\mathbf{\omega}$ will be linear to the
sum of the torques from both tips
\begin{equation}
  \begin{aligned}
    \frac{d \uvec n}{d t} = - \uvec n \times \mathbf{\omega} 
    &\approx - \mu_{\perp} \uvec n \times \langle \vec\tau_\text{head} +
    \vec\tau_\text{tail}\rangle\\
    &=  - \chi\uvec n \times(\uvec n \times \nabla
    \psi)
  \end{aligned}
\end{equation}
where $\mu_{\perp}$ is the rotational mobility and
\begin{equation}
  \chi(\psi) = \mu_{\perp}\ell F'(\psi(\vec r))\langle r_p\sin^2\vartheta\rangle
\end{equation}
which becomes independent of $\psi$ if $F'(\psi) \approx$ const.

In addition to this deterministic reorientation, we should also
account for noise due to thermodynamic fluctuations and fluctuations
in pili attachment, which result in torque fluctuations. For the case
that the pili pulling direction is correlated only on a very short
time scale (i.e. the pili attach according to $P(\vec r_p)$ much
quicker than the microorganism moves), we obtain the stochastic
equation
\begin{equation}
\frac{d \uvec n}{d t} =  -\chi \uvec n \times (\uvec n \times\nabla \psi) +  \uvec n\times \mathbf{\xi}_r
\end{equation}
This equation can be rewritten in terms of the orientation angle
$\varphi$ relative to some axis on the surface by defining $\uvec n =
(\cos \varphi, \sin \varphi)^T$, which gives Eq.~(1c).

\section{Derivation of Equation (1d)}
From the definition of the trail field we have
\begin{equation}
  \label{eq:30a}
  \nabla_{\vec x}\psi(\vec x, t)
  = -\frac{2k}{\pi R^2}\int_0^tdt'[\vec x - \vec r(t')]
  \delta(R^2 - |\vec x - \vec r(t')|^2)\notag
\end{equation}
and with a change of variables $t'\to t - t'$,
\begin{multline}
  \label{eq:40a}
  \partial_{\perp}\psi(\vec r(t), t)
  =-\frac{2k}{\pi R^2}\int_0^tdt'[\vec r(t) - \vec r(t-t')]\cdot\uvec n_{\perp}(t)\\
  \times\delta(R^2 - |\vec r(t) - \vec r(t-t')|^2)
\end{multline}

From the twice iterated integral
\begin{equation}
  \label{eq:9a}
    \vec r(t-\tau) = \vec r(t) + v_0\int_t^{t-\tau}du\uvec n(t)
    + v_0\int_t^{t-\tau}du\int_0^{u}dw\dot{\uvec{n}}(w)\notag
\end{equation}
one finds $[\vec r(t) - \vec r(t-\tau)]^2 = v_0^2\tau^2 + \mathcal
O(\tau^3)$ and
\begin{widetext}
\begin{equation}
  \label{eq:10a}
  [\vec r(t) - \vec r(t-\tau)]\cdot\uvec n_{\perp}(t)
  = - v_0\int_0^{\tau}du\int_0^{u}dw
  \{\chi\uvec e_z\cdot[\uvec n(t-w)\times\nabla\psi(\vec r(t-w), t-w)] + \xi(t-w)\}
  \uvec n_{\perp}(t-w)\cdot\uvec n_{\perp}(t)\notag
\end{equation}  
\end{widetext}
An identical iteration shows $\uvec n_{\perp}(t-w)\cdot\uvec n_{\perp}(t) = 1 + \mathcal O(w)$ and thus
\begin{multline}
  \label{eq:11a}
  [\vec r(t) - \vec r(t-\tau)]\cdot\uvec n_{\perp}(t)\\
  = - v_0\int_0^{\tau}du\int_0^{u}dw[\chi\partial_{\perp}\psi(t-w) +
  \xi(t-w)]
  + \mathcal O(\tau^3)\notag
\end{multline}
Using the above two approximations in Eq.~(\ref{eq:40a}) and performing
one of the integrals yields Eq.~(1d).

\section{Implications on experimental measurements of diffusivities}

\begin{figure}[tb]
  \centering
  \includegraphics{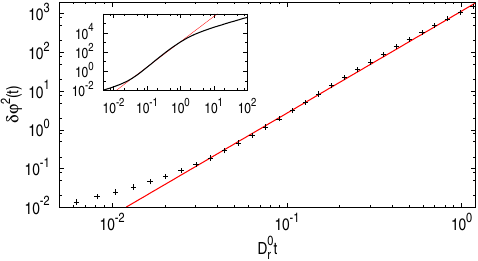}
  \caption{Angular \textsc{MSD} $\delta\varphi^2(t)$ as
    a function of time $t$ for $\Omega\tau=1.94$, $D_r^0\tau=0.1$
    (black crosses) and a power law fit (red line) $\propto t^{2.64}$
    for $D_r^0t\in[0.05, 1]$. The inset shows the same plot over a larger
    time window, showing that the fit \emph{does not} capture the
    asymptotic behavior.}
  \label{fig:dphipower}
\end{figure}

The \textsc{MSD} curve in Fig.~\ref{fig:dphipower} highlights subtleties that
must be considered when interpreting such measurements in terms of a model,
and for extracting model parameters. The long time diffusivity, $D_r$, is
\emph{not} the bare diffusivity $D_r^0$ as determined by the cellular motility
apparatus. Moreover, the asymptotic diffusive regime is reached on very long
time scales only: timescales on the order of hours \footnote{For
  \textit{P. aeruginosa}, $D_r\approx 0.002 \; \rm{s}^{-1}$.}, and that may
well be beyond experimental reach, and even beyond the life time of
microorganisms. In that case, it might be tempting to conclude that the
\textsc{MSD}s show anomalous diffusion asymptotically; as can be seen from
Fig.~\ref{fig:dphipower}, an anomalous exponent $\beta \approx 2.64$ can fit
the data extremely well.

\section{Experimental Methods and Sample Preparation}

\textit{P. aeruginosa} PAO1 \cite{holloway1955genetic} strains $\Delta psl$ --
a strain that does not produce Psl -- as well as
$\Delta$P$_{psl}$/P$_{BAD}$-{\it psl} \cite{ma2006analysis} -- an engineered
Psl-inducible strain -- were used in this study. For the detailed experimental
information such as culture conditions, flow cell assembly and image capture
system, please refer to methods in \cite{zhao+tseng13}. An overnight bacteria
culture in FAB medium \cite{heydorn2000experimental} supplemented with 30mM
glutamate, was diluted and injected into a flow cell. FAB medium with 0.6 mM
glutamate was continuously pumped through the flow chamber using a syringe
pump with a flow rate of 3ml/hour at 30$^{\circ}$C. Different amounts of
arabinose were added into the medium to control the production of
Psl. Bright-field images were taken every 3 seconds by an EMCCD camera on an
Olympus IX81 microscope equipped with Zero Drift autofocus system. The image
size is $67 \times 67 \mu {\rm m}^2$ ($1024 \times 1024 \text{ pixel}^2$). A
typical data set has about 14000$-$20000 frames and contains up to 1,000,000
bacteria images.

\end{document}